# Reflection of light from a uniformly moving mirror

**Aleksandar Gjurchinovski**[*]

*Department of Physics, Faculty of Natural Sciences and Mathematics,*
*Sts. Cyril and Methodius University,*
*P.O.Box 162, 1000 Skopje, Macedonia*

## ABSTRACT

We developed a formula for the law of reflection of a plane-polarized light beam from an inclined flat mirror in uniform rectilinear motion by a direct application of the Huygens-Fresnel principle. Applying the obtained formula and the postulates of special relativity to describe a specific thought experiment, we showed that the moving mirror must be contracted along the direction of its motion by the usual Lorentz factor. The result emphasizes the reality of Lorentz contraction by showing that the contraction is a direct consequence of the first and the second postulate of special relativity, and not a product of the process of relativistic measurement of the length.

**PACS numbers: 03.30.+p, 42.15.Dp, 42.25.Gy**

## I. INTRODUCTION

The experiments involving moving mirrors are among the most interesting experiments encountered in physics. Michelson's setup for measuring the speed of light with a rotating wheel consisted of a mirrored edges, an array of corner mirrors on the Moon's surface for estimating the distance between the Earth and the Moon, the interferometer of Michelson and Morley for detecting the mysterious ether, and the rotating Sagnac interferometer for determining the angular velocity of the Earth are only a few of the setups in which the moving mirrors are playing prominent role. In most of the textbooks in which these experiments are supposed to be carefully explained, it is silently assumed that the ordinary law of reflection of the light is valid - the angles of incidence and reflection are equal. Our goal is to point out that the ordinary law of reflection does not hold in the case when the mirror is moving at constant velocity and to find a correct relation between the incident and the reflected angle.

---

[*] Electronic mail: agjurcin@iunona.pmf.ukim.edu.mk

Reflection of light from a uniformly moving mirror is, by all means, not a new insight [1]. A particular case of the problem was elaborated by Einstein almost a century ago. It was published as a part of his celebrated paper announcing a completely different way of thinking about space and time - the special theory of relativity [2]. Einstein considered oblique incidence of a plane-polarized electromagnetic wave on a perfectly reflecting mirror, whose velocity was directed perpendicularly to its surface. In order to derive the equations for the angle of reflection and the wave characteristics of the reflected light, Einstein implemented Lorentz transformations on the equations describing the reflection in the reference frame where the mirror was at rest.

In the text that follows we will use a different kind of approach. The method we propose is based on the elementary principles of wave optics and the postulates of special relativity. Its simplicity could give us a benefit to bring this apparently forgotten, but important issue, in front of the common undergraduate classroom, as well as to stimulate and increase student's thinking and intuitive understanding of the basic principles of special relativity.

## II. HUYGENS' CONSTRUCTION

To trace the path of an arbitrary light beam through a medium, one usually employs Huygens' construction [3]. It states that every point that belongs to the primary wavefront at some fixed time serves as an elementary source of secondary wavelets which spread out in all directions having the same frequency and velocity as the primary wavefront. The envelope of these wavelets is the wavefront of the light beam at a later time. If the medium through which light propagates is made of an optically isotropic substance, then the light ray can be constructed as a line normal to every subsequent wavefront at all times.

We will implement an atomic version of the Huygens-Fresnel principle to describe reflection of light by an inclined plane mirror moving at constant velocity $v$ in vacuum. The situation is shown in Fig. 1. By $\varphi$ we denote the inclination angle of the moving mirror that defines the slope of the mirror's surface with respect to the negative direction of the $x$ - axis.

Let 1 and 2 be the boundary rays of the incident plane-polarized light beam, and the distance $\overline{AB}$ the wavefront of the incoming light at some time $t_0$. The atoms at the point $A$ are disturbed by the incident wavefront and they start to radiate a wavelet. The disturbance process continues with the following atoms along the surface of the moving mirror, and it stops at the time $t$ when the wavefront strikes the point $D$. The atoms at the final point $D$ become a source of wavelets $(t - t_0)$ seconds after the initial disturbance of the atoms at the point $A$. In the case when the mirror is stationary, an elementary wavefront emitted from a source on the mirror's surface would be an expanding sphere, whose radius at the instant $\tau$ would equal $c\tau$. By $\tau$ we denote the amount of time expired from the beginning of the emission, and $c$ is the speed of light in vacuum. In the present case, when the mirror is in uniform rectilinear motion, the elementary wavefront will remain a sphere, expanding equally in all directions at the same constant speed $c$ [4]. The last statement is a direct consequence of the second postulate of special relativity - the speed of light in vacuum is a universal constant, and its value $c$ ($\approx 3 \cdot 10^8 \, m/s$) is independent of the motion of the source. It can be further clarified with the fact that the equation that describes the evolution of an elementary wavefront in vacuum is invariant under Lorentz transformations [5]. Consequently, at the moment when the incident wavefront $\overline{AB}$ will reach the point $D$, the elementary wavefront emitted from the source at $A$ will represent a sphere with radius $\overline{AC} = c(t - t_0)$.

The motion of the mirror causes the elementary sources of the secondary wavelets to lie along the straight line connecting the points $A$ and $D$. The envelope of all these elementary wavelets is the distance $\overline{CD}$, which is, actually, the reflected wavefront, and 1′ and 2′ are its boundary rays (see Fig. 1). It is obvious that the optical disturbance of every point along $\overline{CD}$ has a constant phase. We denote by $\alpha$ the angle of incidence of the wavefront towards the normal $n$ of the mirror's surface, and by $\beta$ the angle of its reflection. If $v = 0$, then, by implementing certain triangle similarities, one can prove the well known relation $\alpha = \beta$ [6]. However, the ordinary law of reflection would not be valid when the mirror is moving.

At this point, we would like to give a brief description of the shifting phenomenon, whose appearance is a result of the mirror's motion and the finite profile of the incident light (Fig. 2). If we take the incident light beam to emerge from a stationary source, then the wavefront $\overline{A'B'}$, which is incident on the moving mirror subsequently to the initial wavefront $\overline{AB}$, will be reflected at the same angle $\beta$ as the wavefront $\overline{AB}$ but from slightly different points on the mirror's surface. Hence, the boundary rays 1″ and 2″ of the reflected wavefront $\overline{C'D'}$ will not coincide with the boundary rays 1′ and 2′ of the initially reflected wavefront $\overline{CD}$, but they will be shifted up to an infinitesimal distance with respect to 1′ and 2′ (see Fig. 2). We hereby note that the reflected wavefronts will be parallel to each other and they will have equal widths. The shifting process continues with every consecutive wavefront until the moving mirror "escapes" the incident light. If a flat screen is placed against the reflected light, then, as a consequence of this shifting process, the projected profile of the reflected beam will not be fixed, but it will be moving at constant velocity (see Fig. 3).

In the following we will concentrate our attention on the wavefront $\overline{AB}$ and its reflected analogue $\overline{CD}$. From Fig. 1, we have

$$\sin \alpha = \frac{\overline{BD} + \overline{DG}}{\overline{AG}}, \tag{1}$$

$$\sin \beta = \frac{\overline{AC} - \overline{AF}}{\overline{AG} - \overline{EF}}. \tag{2}$$

We have taken into account that $\overline{ED} = \overline{AG}$. From the discussion associated with Fig. 1, we have

$$\overline{AC} = \overline{BD} = c(t - t_0).\tag{3}$$

Figure 4 is an enlargement of the area around the point $A$. Observe that $d = \overline{AO} = v(t - t_0)\sin\varphi$ is the shortest distance between the positions of the moving mirror at the times $t_0$ and $t$. Thus, the following equations are in order

$$\overline{DG} = \overline{AE} = \frac{\overline{AO}}{\cos\alpha} = \frac{v(t - t_0)\sin\varphi}{\cos\alpha},\tag{4}$$

$$\overline{AF} = \frac{\overline{AO}}{\cos\beta} = \frac{v(t - t_0)\sin\varphi}{\cos\beta}.\tag{5}$$

From the triangles $AEO$ and $AFO$, we have $\overline{EO} = \overline{AO}\tan\alpha$ and $\overline{OF} = \overline{AO}\tan\beta$. But $\overline{EF} = \overline{EO} + \overline{OF}$, which leads to

$$\overline{EF} = v(t - t_0)\sin\varphi\,(\tan\alpha + \tan\beta).\tag{6}$$

We substitute Eqs. (3), (4), (5) and (6) into Eqs. (1) and (2) to obtain

$$\sin\alpha = \frac{c + v\dfrac{\sin\varphi}{\cos\alpha}}{\dfrac{\overline{AG}}{(t - t_0)}},\tag{7}$$

$$\sin\beta = \frac{c - v\dfrac{\sin\varphi}{\cos\beta}}{\dfrac{\overline{AG}}{(t - t_0)} - v\sin\varphi\,(\tan\alpha + \tan\beta)}.\tag{8}$$

A straightforward elimination of the term $\overline{AG}/(t - t_0)$ from Eqs. (7) and (8) and a little algebra would result in a relation that links the incident angle $\alpha$ and the reflected angle $\beta$ in the following manner

$$\sin\alpha - \sin\beta = \frac{v}{c}\sin\varphi\sin(\alpha + \beta).\tag{9}$$

Equation (9) is the law of reflection of light from an inclined flat mirror in uniform rectilinear motion. Obviously, when the mirror is at rest ($v = 0$) or when its inclination angle is zero ($\varphi = 0$), the angles of incidence and reflection would be equal. Evidently, if $\alpha = 0$, then $\beta = 0$ for any $\varphi$ and $v$. In the case when the angle of incidence differs from zero ($\alpha \neq 0$), Eq. (9) can be rewritten in a more compact shape

$$\frac{\sin\alpha - \sin\beta}{\sin(\alpha + \beta)} = \frac{v}{c}\sin\varphi \ . \tag{10}$$

By following the previous procedure, one can show that the law of reflection of the light when the mirror is moving in the opposite direction to the one shown in Fig. 1 is

$$\frac{\sin\alpha - \sin\beta}{\sin(\alpha + \beta)} = -\frac{v}{c}\sin\varphi \ . \tag{11}$$

The last equation can be also obtained from Eq. (10) by putting $-v$ instead of $v$, or, equivalently, $-\varphi$ instead of $\varphi$.

Although at first sight it would appear that Eqs. (9), (10) and (11) are transcendental equations, careful investigation shows that this is not so. Quite a contrary, one can derive a formula for the angle of reflection (incidence) directly in terms of the angle of incidence (reflection), the inclination angle $\varphi$ and the velocity $v$ of the moving mirror. The procedure is described in Problem 3 of the Appendix for a special arrangement, but it can be also implemented for the general situation when the moving mirror makes an arbitrary angle $\varphi$ with respect to the negative direction of its velocity vector. The reader is encouraged to show that, for example, Eq. (10) leads to the following formula for the reflected angle $\beta$

$$\cos\beta = \frac{2\dfrac{v}{c}\sin\varphi + \left(1 + \dfrac{v^2}{c^2}\sin^2\varphi\right)\cos\alpha}{1 + 2\dfrac{v}{c}\sin\varphi\cos\alpha + \dfrac{v^2}{c^2}\sin^2\varphi} \ . \tag{12}$$

However, in order to make the derivations in the subsequent sections as simple as possible, we will make use of the law of reflection given in the forms of Eqs. (9), (10) and (11).

We further note that the obvious asymmetric treatment of the angles of incidence and reflection in Eq. (10) is followed by an important consequence. Namely, if the reflected light ray becomes incident, it will not be reflected at an angle $\alpha$, but at some different angle $\delta$ which is a solution of the equation

$$\frac{\sin \beta - \sin \delta}{\sin(\beta + \delta)} = \frac{v}{c} \sin \varphi .$$ (13)

In other words, the principle of reversibility of the light rays doesn't hold if the light is reflected by a moving mirror.

To conclude this section, we would like to emphasize the following. It is obvious that the inclination angle $\varphi$ is a substantial constituent of the law of reflection given by Eq. (9). However, the derivation of Eq. (9) was based on the second postulate of special relativity. Nevertheless, the angle $\varphi$ is a real, physical entity, which, by itself, has nothing to do with relativity. The value of $\varphi$ is neither a result of an act of measurement, nor a quantity obtained upon perception [7]. In order to make this point further clear, we urge the reader to recall that while applying the Huygens-Fresnel principle in the derivation of Eq. (9), we assumed that the surface of the moving mirror was made of atoms, each of which was moving at constant velocity $v$ as the mirror, and each of which would radiate secondary wavelets if disturbed by the incident light. The inclination angle $\varphi$ was introduced in order to describe the moving plane on which these atoms are situated. Thus, the angle $\varphi$ is the physical angle that the mirror makes at any instant with respect to the negative direction of its velocity vector. We will utilize this fact in the following section in order to derive an important inherent property of an inclined flat mirror in uniform rectilinear motion.

## III. EINSTEIN'S CAT EXPERIMENT

Consider the experimental setup shown in Fig. 5. The light beam emanating vertically downward from the light source at $A$ is reflected by a flat mirror at the point $B$. The mirror is inclined to $\pi/4$ radians with respect to the negative $x$ - axis. The reflected beam hits the point $C$, which belongs to a chamber consisting of two rooms separated by a surface with electronically controlled permeability. By hitting the point $C$, the beam activates a sophisticated life-supporting mechanism, which prevents the poisonous gas from entering the room where the cat is located by acting on the permeability of the protecting surface.

Now, what happens when the experiment is observed from a frame in which the whole setup is moving to the right at velocity $v$? One can follow the traditional scheme as in most of the textbooks and say that Fig. 6 shows the path of the light beam in the case when the setup is in uniform rectilinear motion [8]. However, this is not a correct answer, as we shall see in the following. We offer an explanation by using the principles of special relativity.

While examining Fig. 6, the reader must be cautious and try to recognize that the bold line $AB'C'$ is not the path of the light beam itself, but a path traversed by a single wavefront, the one emitted at the instant of time when the source of the light beam was positioned at the point $A$. All the wavefronts following this one will, of course, be emitted by the same source, but from different points in space. Due to the motion of the source, these points will be located to the right of the point $A$. The plane of every single wavefront on its journey from the moving source to the moving mirror will be perpendicular to the wavefront's trajectory, and the velocity of the wavefront along its trajectory will be constant and equal to $c$ [9]. The later is, actually, Einstein's second postulate at work. The mirror is obviously having the same velocity $v$ as the source, which means that at any point of time along the course of their motion, the moving source and the moving mirror will lie on the same vertical line.

By looking at Fig. 6 we see that the distance traversed by the initially emitted wavefront from its source at $A$ to the mirror at $B'$ is $\overline{AB'} = ct$, where $t$ is the time required for the wavefront to cover the distance $\overline{AB'}$ at velocity $c$. On another hand, the distance traveled by the light source, or equivalently, by the mirror, for the same amount of time, is $\overline{BB'} = \overline{AA'} = vt$. Consequently, from the triangle $AA'B'$ we have

$$\sin\theta = \frac{v}{c}, \qquad\qquad (14)$$

which is the well known formula for the aberration of light [10]. The path from the moving source to the moving mirror traversed by every subsequently emitted wavefront will be parallel to the path $\overline{AB'}$ of the initially emitted wavefront. At every instant of time, all the wavefronts propagating from the moving source to the moving mirror will be lined up along the vertical line connecting the moving source and the moving mirror (see Fig. 7). They will be reflected by the moving mirror from the same range of points (i.e. atoms) along its surface as the initially reflected wavefront. Therefore, if the experiment in Fig. 5 is observed from a reference frame traveling to the left at constant velocity $v$, it would appear that the light beam is advancing at velocity $c\sqrt{1 - v^2/c^2}$ along the vertical line connecting the moving source and the moving mirror, while, at the same time, the whole setup (including the light beam!) is moving at velocity $v$ to the right (Fig. 8).

According to Einstein's first postulate (the principle of relativity), all the inertial frames are equivalent (i.e. there are no preferred inertial reference frames), and the laws of physics are identical in all of them. In this case, it means that if the light is hitting the switch of the chamber in the reference frame where the setup is at rest (Fig. 5), then the light must hit the switch of the chamber in every other inertial reference frame, regardless of the direction of motion of the setup. In simple words, if the cat is alive in one inertial reference frame, then the cat must stay alive in every other inertial reference frame. The previous considerations require that every single wavefront must be reflected by the moving mirror, and, in the present case (Fig. 6), must follow the horizontal in order to hit the switch of the chamber. Someone might argue the possibility that the light beam will hit the mirror at some point other than $B'$, or that the light is somehow hitting $C'$ directly, and not by reflection from the moving mirror. However, we can modify the setup in Fig. 5 by making the point $B$ to be an on-off switch of another life-supporting mechanism belonging to another cat in a chamber. If we assume that the person is right, apparently, in the frame of reference where the setup is moving (Fig. 6), the beam will miss the switch at the point $B'$, the second life-supporting mechanism will not be activated, and the cat will be dead. By *reductio ad apsurdum*, we came to the conclusion that the light beam must hit the moving mirror at $B'$.

Now, when we are convinced that the situation shown in Fig. 6 is a correct one, but it stands for an individual wavefront (in this case, the one emitted from the source at $A$), we will verify the consistency of Eq. (9). We will assume that Eq. (9) correctly describes the reflection of the wavefronts of the incident light from the surface of the moving mirror. Furthermore, we will also assume that the inclination angle $\varphi$ of the moving mirror may differ from the inclination angle of the stationary mirror (in our case, $\pi/4$ radians). From Fig. 6 we express the incident angle $\alpha$ and the reflected angle $\beta$ as

$$\alpha = \theta + \varphi \,, \tag{15}$$

$$\beta = \pi/2 - \varphi \,. \tag{16}$$

By replacing Eqs. (15) and (16) into Eq. (10), we have

$$\frac{\sin(\theta + \varphi) - \sin(\pi/2 - \varphi)}{\sin(\theta + \varphi + \pi/2 - \varphi)} = \frac{v}{c} \sin \varphi \,. \tag{17}$$

After some rearrangements, and taking into account Eq. (14), we get

$$\tan \varphi = \frac{1}{\cos \theta} = \frac{1}{\sqrt{1 - \sin^2 \theta}} = \frac{1}{\sqrt{1 - \dfrac{v^2}{c^2}}} \,. \tag{18}$$

The result shows that $\tan \varphi \neq 1$, which means that our assumption was in order and that the moving mirror really makes a different inclination angle than the same mirror at stationary condition. Here we emphasize the fact that the lateral dimensions of an object do not change when the object is in uniform rectilinear motion [11]. The last assertion implies that the alteration of the inclination angle of the mirror due to its uniform motion can be caused only by alteration of the mirror's dimensions parallel to the direction of its motion. With this in mind, we can express the inclination angle of the moving mirror as

$$\tan \varphi = \frac{l_0}{l} \,, \tag{19}$$

where by $l_0$ we denote the projection of the mirror's length on the axis perpendicular to its motion, and by $l$ the projection of the mirror's length on the direction of its motion. Then, by applying Eq. (19) into Eq. (18), we have

$$l = l_0 \sqrt{1 - \frac{v^2}{c^2}}, \qquad (20)$$

which is the well known formula in special relativity, commonly abbreviated as Lorentz contraction of the length. Equation (20) states that an inclined flat mirror moving at constant velocity $v$ will be Lorentz contracted along the axis of its motion.

Let us explore an identical setup, but now we will exchange the positions of the light source $A$ and the chamber $C$. The setup with respect to the reference frame where the mirror is at rest is shown in Fig. 9. Following the same arguments as in the previous case, one can demonstrate that Fig. 10 shows the whole situation (for an individual wavefront !) observed from a frame in which the setup is in motion. In this case,

$$\alpha = \pi/2 - \varphi, \qquad (21)$$

$$\beta = \varphi - \theta. \qquad (22)$$

By substitution of Eqs. (21) and (22) into Eq. (9), and taking into account Eqs. (14) and (19), we get

$$l = l_0 \sqrt{1 - \frac{v^2}{c^2}}. \qquad (20)$$

Obviously, the mirror will be shortened along the line of its motion in the same manner as in the previous case.

Assuming the correctness of Eq. (9) and using Einstein's postulates, we were able to derive the relativistic contraction formula for an inclined flat mirror in uniform rectilinear motion. The result is in accordance with the predictions of special relativity, which means that Eq. (9) correctly describes the propagation of the wavefront reflected by the moving mirror. If we assumed that instead of Eq. (9) the usual law of reflection is in order (that is, the angle of incidence equals the reflected angle), we would arrive at peculiar results, not just contradictory to the predictions of special relativity, but also against the common sense (see Problem 1 in Appendix).

If the ray optics of the original Einstein's cat experiment shown in Fig. 5 is somehow modified, for example, the inclination angle of the stationary mirror is set to be different than $\pi/4$ while the position of the source and the angle of the emerging light are adjusted in such a manner that the path of the reflected beam remains unchanged, i.e. it follows the horizontal in order to hit the switch at the point *C*, then the effect of relativistic contraction of the moving mirror along its velocity vector would still be in accordance with Eq. (20). Different versions of Einstein's cat experiment can be used as homework problems.

## IV. SUMMARY AND DISCUSSION

The first thing we would like to emphasize is that the formula for the law of reflection of light from a uniformly moving mirror expressed by Eq. (9) is a consequence of the constant light speed postulate. While deriving Eq. (9), we argued that as a result of Einstein's second postulate, the shape of every single elementary wavefront originating from a source on the moving mirror's surface will remain unchanged with respect to the shape of the wavefront when the mirror is stationary. It will represent a sphere, expanding equally in all directions at constant speed *c*. The rest of the derivation is a classical usage of the standard Huygens' construction.

By making use of the atomic version of the Huygens-Fresnel principle, we showed that the angle of reflection of the light depends on the angle of incidence, the inclination angle $\varphi$ and the velocity *v* of the moving mirror. Applying the obtained formula and the principles of relativity to describe a specific thought experiment, we arrived at the result that the inclination angle of the moving mirror would differ from its inclination angle in a stationary condition. As we previously stressed that the inclination angle $\varphi$ is the physical angle that the mirror makes at any instant with respect to the negative direction of its velocity vector, we may conclude that this tilt of the mirror due to its motion is a real effect. We further showed that this tilting phenomenon is actually a consequence of the fact that the physical length of the moving mirror in the direction of its motion is less than the physical length of the same mirror at rest by the usual Lorentz factor.

In the special theory of relativity the length of an object in a given inertial frame is defined as the distance between any two simultaneous events that occur at the object's ends. According to this definition and the Lorentz transformation formulas, an object moving at constant velocity will be Lorentz contracted along the direction of its motion. A common interpretation of the effect of Lorentz contraction states that the relativistic contraction is not a real, physical contraction, but an apparent, artificial phenomenon, whose presence is solely due to the process of relativistic measurement of the length. By relativistic measurement we mean a measurement performed by a stationary observer with a stationary measuring equipment (rulers, clocks, etc.) and by involving the notion of simultaneity [12]. Hence, a uniformly moving object will be measured to be Lorentz contracted along the direction of its motion, while its physical length will remain the same no matter what reference frame one uses to describe it [13]. The effect of Lorentz contraction does not exist apart from the measuring process. It only occurs when the length of an object is measured in a relativistic sense. However, in the present paper we were able to approach the problem in a specific way and to show that Lorentz contraction is a real effect - an inherent physical property of a mirror (and, therefore, any object) in uniform rectilinear motion [14]. As a physical property of an object in motion, the phenomenon of Lorentz contraction is not a product of the process of relativistic measurement of the object's length.

While describing Einstein's cat experiment, we argued that some important points are usually omitted or not correctly tackled in the similar exposures. We showed that the bold line depicted in Fig. 6 is not the path of the light beam itself, but a path of a singular wavefront emitted at the instant of time when the light source was positioned at the point $A$. The actual advancement of the light beam from the source to the mirror, with respect to the frame of reference in which the setup is in motion, is described in Fig. 8. There should be no confusion with the statement that the light beam is advancing at a speed $c\sqrt{1-v^2/c^2}$ along the vertical line between the source and the mirror. This statement does not contradict Einstein's second postulate. As we pointed out, at the same time the beam as a whole is moving at velocity $v$ together with the rest of the setup, which will cause every point (i.e. wavefront) along the beam to possess a net constant speed $c$, whose direction is determined by the aberration angle $\theta$.

If the previous considerations are not taken into account, one can be easily surrounded by a paradoxical situation. Suppose that the path in Fig. 6 is the real path taken by the light beam. Then, for example, for the concrete case $v = 0.85c$, the angle of incidence $\alpha$ will be greater than $\pi/2$ even without taking relativistic contraction into account (detailed calculations with and without Lorentz contraction are left to the student as an exercise). Hence, the light beam will strike the back side of the mirror, and, therefore, it will never reach the switch at $C'$. Consequently, and against Einstein's first postulate, the cat will be dead.

## V. CONCLUSION

Reflection of light from a uniformly moving mirror is an issue commonly omitted by the standard textbooks on optics. Surprisingly, the topic appears to be unexplored even by the advanced treatments of relativistic electrodynamics, with certain exceptions where some specific arrangements of the problem are considered.

The problem dates back to Einstein's monumental work on special relativity. Einstein correctly solved the problem in the framework of the newly developed theory. We have shown that the subject can be approached alternatively, in a way accessible for undergraduate, or even high school audience. We developed a formula for the law of reflection of light from an inclined flat mirror in uniform rectilinear motion by using an elementary ray-tracing method, and we used the formula to describe a couple of specific thought experiments. The result of the Einstein's cat experiment emphasizes the reality of Lorentz contraction by showing that the contraction is a direct consequence of the first and the second postulate of special relativity, and not a product of the process of relativistic measurement of the length.

Furthermore, we showed that certain conceptions concerning the optical setups observed from a moving frame of reference are not fully understood or described in the textbooks. These conceptions must be carefully reconsidered if the teacher or student is about to give a correct description of an optical experiment involving inclined plane mirrors in a frame of reference where the setup is in motion à la special relativity.

**APPENDIX - PROBLEMS**

*Problem 1.* **Failure of the classical law of reflection**

By using the postulates of special relativity, show that for the situations depicted in Figs. 6 and 10 the angles of incidence and reflection cannot be equal.

*Solution.* Assume that for the setup shown in Fig. 6, the angles of incidence and reflection are equal, that is, $\alpha = \beta$. Then, by equalizing the right hand sides of Eqs. (15) and (16), we get $\theta = \pi/2 - 2\varphi$. By implementing Eq. (14), we have

$$\sin \theta = \sin(\pi/2 - 2\varphi) = \cos 2\varphi = \cos^2 \varphi - \sin^2 \varphi = \frac{1 - \tan^2 \varphi}{1 + \tan^2 \varphi} = \frac{v}{c}. \qquad (A1)$$

Applying Eq. (19), we may conclude that the mirror will be elongated in the direction of $v$ according to the formula

$$l = l_0 \sqrt{\frac{1 + v/c}{1 - v/c}}. \qquad (A2)$$

Repeating the whole procedure for the setup in Fig. 10 under the same assumption $\alpha = \beta$, we arrive at the result that the mirror will be shortened in agreement with

$$l = l_0 \sqrt{\frac{1 - v/c}{1 + v/c}}. \qquad (A3)$$

One might speculate that the moving mirror is somehow adjusting itself towards the light beam, and its "rotation" depends on the angle of incidence of the incoming wavefront. Now, let us investigate the setup depicted in Fig. 11. This setup is a kind of a combination of the setups shown in Figs. 5 and 9. We have two parallel light beams emerging from different sources and hitting two different chambers after being reflected from a stationary mirror. The position of the second source $A_2$ is adjusted horizontally towards the mirror in such a way that when the whole experiment is observed from the reference frame where the setup is moving at constant velocity $v$ to the right (Fig. 12), after a simultaneous flash of the light sources $A_1$ and $A_2$, the two initially emitted wavefronts will hit the mirror at the same time. In the frame of reference where the setup is in motion, at the instant of time when these two wavefronts will hit the mirror, according to Eqs. (A2) and (A3), the mirror will be elongated and shortened simultaneously. The resulting contradiction proves the assumption $\alpha = \beta$ to be wrong.

*Problem 2*. **Michelson-Morley experiment**

Prove that in the Michelson-Morley experiment the light rays leaving the interferometer will meet on parallel paths [15, 16, 17, 18].

*Solution*. Figure 13 is a schematic description of the Michelson-Morley interferometer in uniform rectilinear motion at constant velocity $v$ to the right. The incident light beam is divided into two beams by a half-silvered mirror at the point $A$. After traversing different paths in the apparatus, the wavefronts of these two beams are recombined at the point $A''$. Before recombining at $A''$, the wavefront of the second (horizontal) light beam was previously reflected by a third mirror not shown in the figure.

The law of reflection of the wavefront reflected at the point $A$ states

$$\frac{\sin\alpha - \sin\beta}{\sin(\alpha+\beta)} = -\frac{v}{c}\sin\varphi \ . \tag{A4}$$

By looking at Fig. 13 we see that the following relations are valid

$$\alpha = \pi/2 - \varphi \ , \tag{A5}$$

$$\beta = \pi/2 + \varphi - \theta_1 \ . \tag{A6}$$

Substitution of Eqs. (A5) and (A6) into Eq. (A4) would yield a result which can be further simplified to give

$$\theta_1 = 2\arctan\sqrt{\frac{1 - v/c}{1 + v/c}} \ . \tag{A7}$$

The wavefront reflected at $A$ strikes the second mirror at $B'$. In this case, the angles of incidence and reflection are equal. The fact follows from Eq. (10) if we take into account that the inclination angle of the mirror at the point $B'$ equals $\pi$ radians.

For the wavefront reflected at the point $A''$ we have

$$\frac{\sin\delta - \sin\omega}{\sin(\delta + \omega)} = \frac{v}{c}\sin\varphi,$$ (A8)

$$\delta = \pi/2 - \varphi,$$ (A9)

$$\omega = \theta_2 - \pi/2 + \varphi.$$ (A10)

Putting Eqs. (A9) and (A10) into Eq. (A8), and after some algebra, we obtain

$$\theta_2 = 2\arctan\sqrt{\frac{1 - v/c}{1 + v/c}}.$$ (A11)

We came to the conclusion that $\theta_1 = \theta_2$, which means that the light rays emerging the apparatus will meet on exactly parallel paths. As a homework problem, the student can try to show that Eq. (A11), or, equivalently, Eq. (A7), can be simplified toward

$$\tan\theta_{1,2} = \frac{c}{v}\sqrt{1 - \frac{v^2}{c^2}}.$$ (A12)

The result is identical to the one obtained by Schumacher [15].

*Problem 3*. **Einstein's mirror**

In his monumental work *On the electrodynamics of moving bodies* [2], Einstein derived the law of reflection of a plane-polarized electromagnetic wave from a flat mirror moving at constant velocity $v$ in vacuum (see Fig. 14). By applying Lorentz transformations on the equations derived in the reference frame where the mirror was at rest, he arrived at the formula

$$\cos\beta = \frac{\left(1 + \dfrac{v^2}{c^2}\right)\cos\alpha - 2\dfrac{v}{c}}{1 - 2\dfrac{v}{c}\cos\alpha + \dfrac{v^2}{c^2}}.$$ (A13)

Show that this result can be also obtained from Eq. (9).

*Solution*. In the case under consideration, after substituting $\varphi = \pi/2$ and putting $-v$ instead of $v$, Eq. (9) looks as follows

$$\sin \alpha - \sin \beta = -\frac{v}{c} \sin(\alpha + \beta) \,. \tag{A14}$$

We use the basic trigonometric identities and some algebra to obtain

$$\left(1 + \frac{v}{c} \cos \beta\right) \sin \alpha = \left(1 - \frac{v}{c} \cos \alpha\right) \sin \beta \,. \tag{A15}$$

By taking the square of the last equation and rearranging the terms, we arrive at the quadratic equation in $\cos\beta$,

$$\left(1 - 2\frac{v}{c} \cos \alpha + \frac{v^2}{c^2}\right) \cos^2 \beta + 2\frac{v}{c}(1 - \cos^2 \alpha) \cos \beta + 2\frac{v}{c} \cos \alpha - \left(1 + \frac{v^2}{c^2}\right) \cos^2 \alpha = 0 \,, \tag{A16}$$

whose solutions are

$$(\cos \beta)_1 = \frac{2\frac{v}{c} \cos^2 \alpha - \left(1 + \frac{v^2}{c^2}\right) \cos \alpha}{1 - 2\frac{v}{c} \cos \alpha + \frac{v^2}{c^2}} \,, \tag{A17}$$

and

$$(\cos \beta)_2 = \frac{-2\frac{v}{c} + \left(1 + \frac{v^2}{c^2}\right) \cos \alpha}{1 - 2\frac{v}{c} \cos \alpha + \frac{v^2}{c^2}} \,. \tag{A18}$$

In the reference frame where the mirror is stationary ($v = 0$), the angles of incidence and reflection must be equal. But, if $\alpha = \beta$, then $\cos\alpha = \cos\beta$. By substituting $v = 0$ in Eqs. (A17) and (A18), we get

$$(\cos \beta)_1 = -\cos \alpha \,, \tag{A19}$$

and

$$(\cos \beta)_2 = \cos \alpha \,. \tag{A20}$$

It follows that the only physically significant solution is

$$(\cos \beta)_2 = \frac{-2\dfrac{v}{c} + \left(1 + \dfrac{v^2}{c^2}\right)\cos\alpha}{1 - 2\dfrac{v}{c}\cos\alpha + \dfrac{v^2}{c^2}}, \qquad (A18)$$

which exactly matches the formula (A13) derived by Einstein in an alternative way.

**ACKNOWLEDGMENTS**

I would like to take this opportunity and to acknowledge the anonymous referee for a thorough revision process and for a set of very useful, constructive and thought-provoking comments and suggestions, without which the key idea of this article would have not been written. I also thank Prof. Hendrik Ferdinande (University of Ghent) for his continuous support of my work.

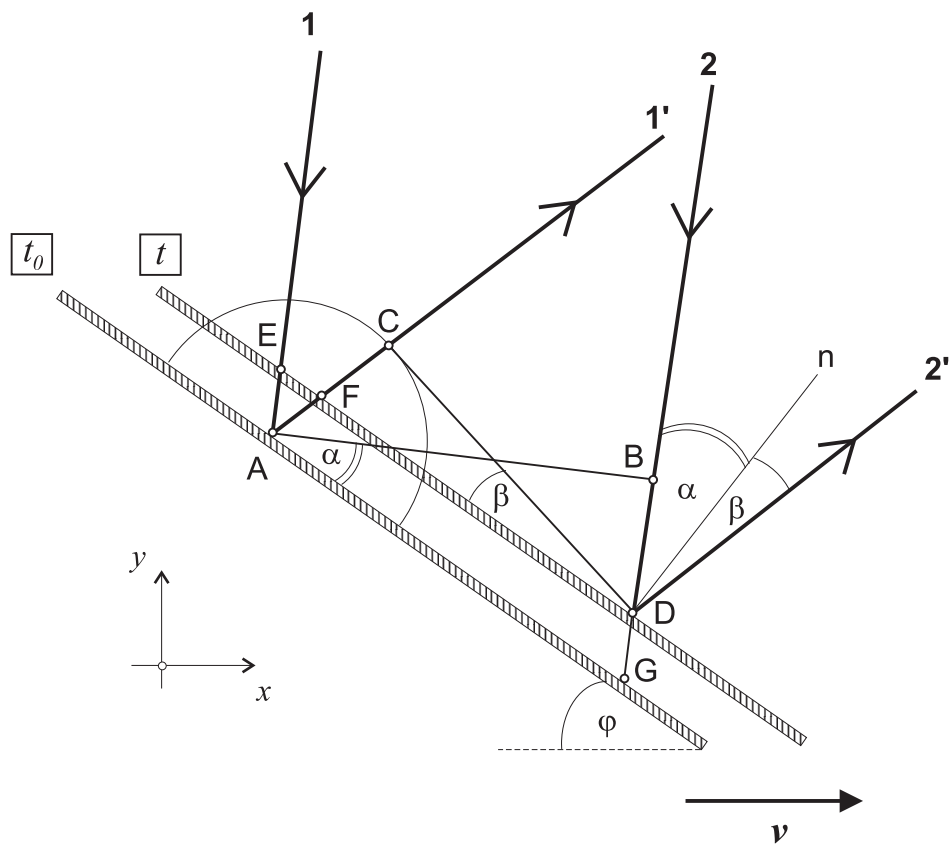

GjurchinovskiFig1

**Fig. 1.** Huygens' construction of the reflected wavefront when the mirror is moving at constant velocity *v* along the positive direction of the *x* - axis.

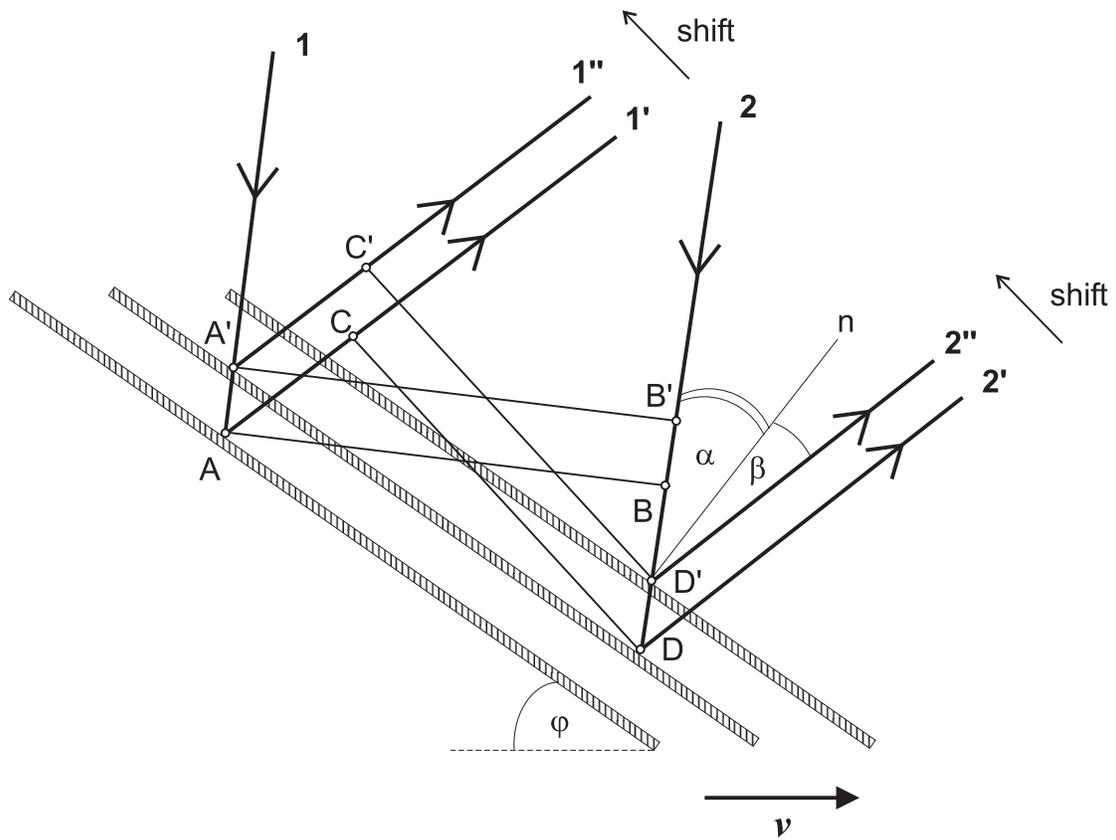

GjurchinovskiFig2

**Fig. 2.** Schematic description of the shifting phenomenon due to the motion of the mirror. The distances between the consecutive wavefronts and between the boundary rays of the reflected wavefronts are exaggerated for convenience.

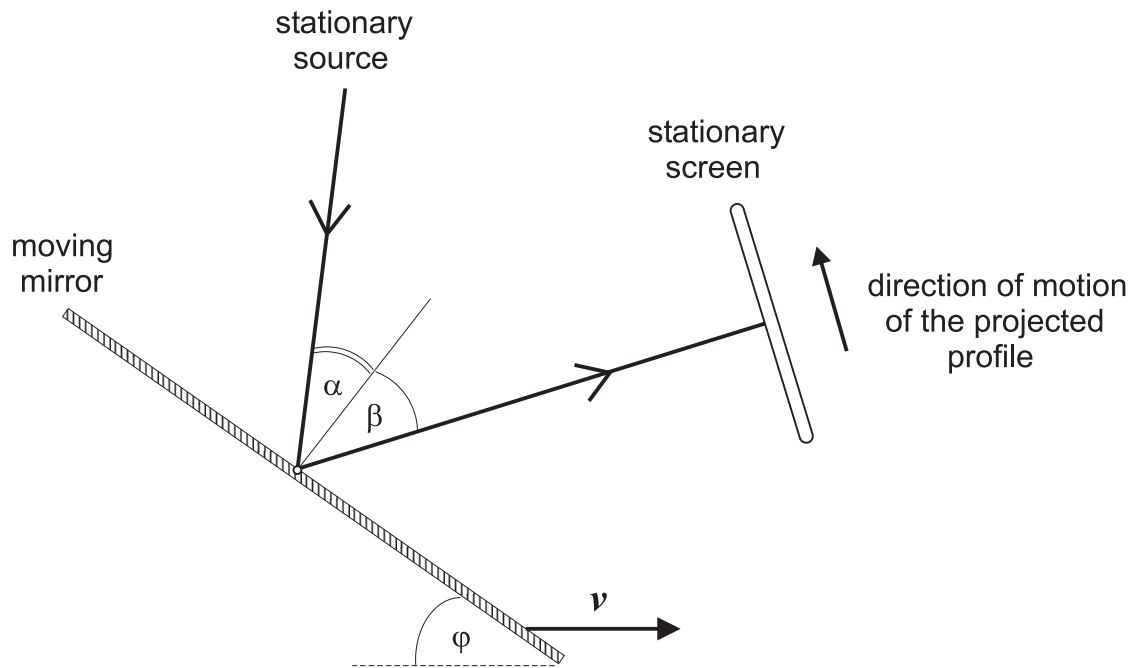

GjurchinovskiFig3

**Fig. 3.** As a consequence of the motion of the mirror, the profile of the reflected beam projected on a stationary screen will be moving at constant velocity.

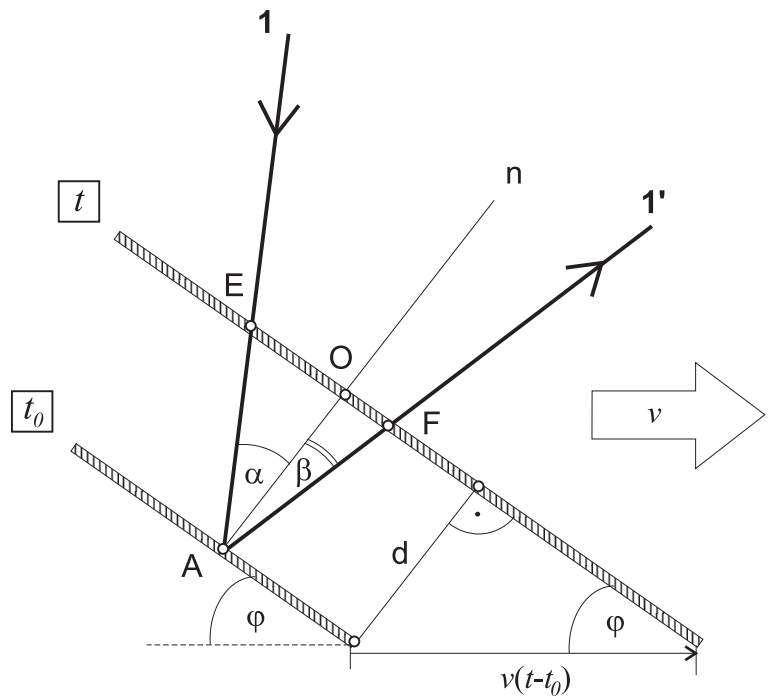

GjurchinovskiFig4

**Fig. 4.** Enlargement of the area around the initially disturbed point *A* on the moving mirror.

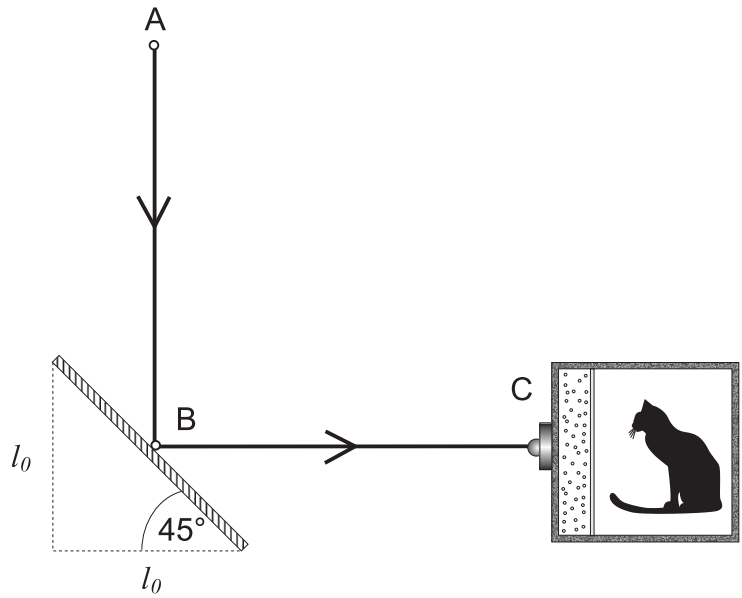

GjurchinovskiFig5

**Fig. 5.** Einstein's cat experiment.

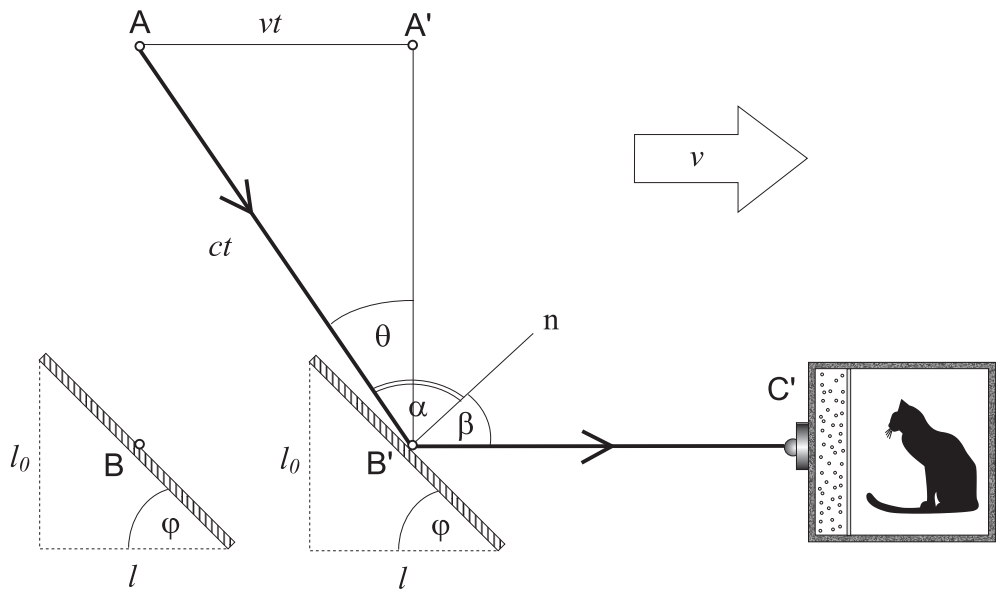

GjurchinovskiFig6

**Fig. 6.** The experiment shown in Fig. 5, observed in a reference frame where the setup is in uniform rectilinear motion. It must be stressed that the bold line $AB'C'$ is not the path of the light beam itself, but a path of a singular wavefront emitted at the instant of time when the light source was positioned at the point $A$.

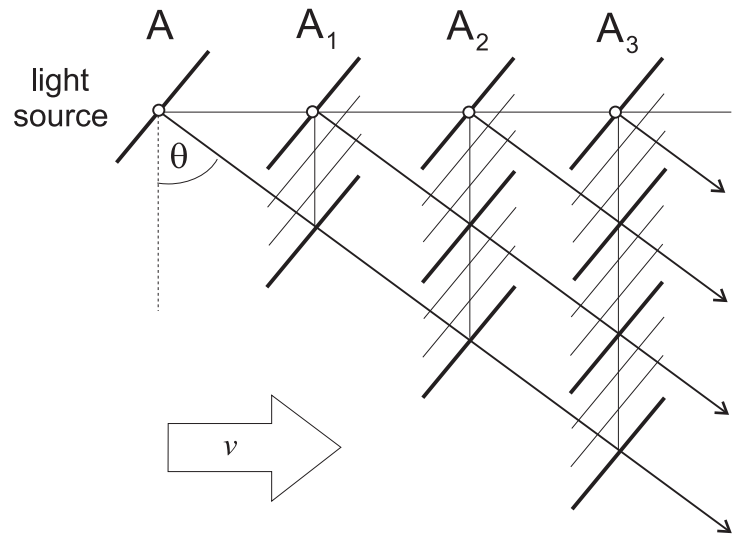

GjurchinovskiFig7

**Fig. 7.** Several consecutive snapshots of propagation of the wavefronts from the moving source to the moving mirror. The snapshots were taken at times when the light source was positioned at the points $A$, $A_1$, $A_2$ and $A_3$, consequently, along the line of its motion. At every instant of time, all the wavefronts will be lined up along the vertical line connecting the moving source and the moving mirror. Observe that the plane of every propagating wavefront will be perpendicular to the wavefront's path, but not to the vertical line along which the light beam advances.

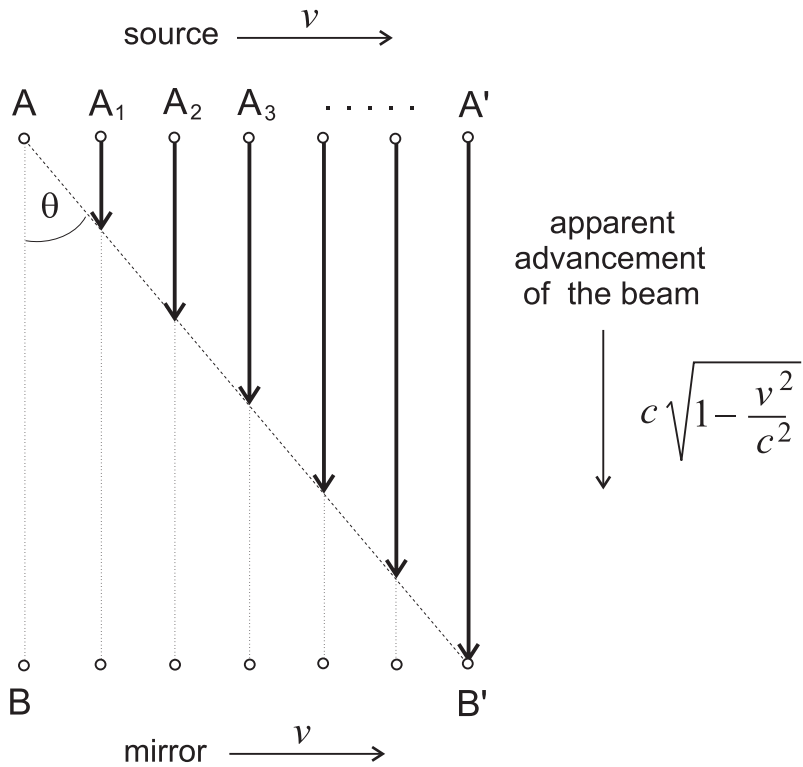

GjurchinovskiFig8

**Fig. 8.** Several consecutive snapshots of the advancement of the light beam from the source to the mirror, observed from a reference frame traveling to the left at constant velocity *v*. While the light beam is advancing at a speed $c\sqrt{1-v^2/c^2}$ along the vertical line between the source and the mirror, the beam as a whole is moving at velocity *v* together with the rest of the setup.

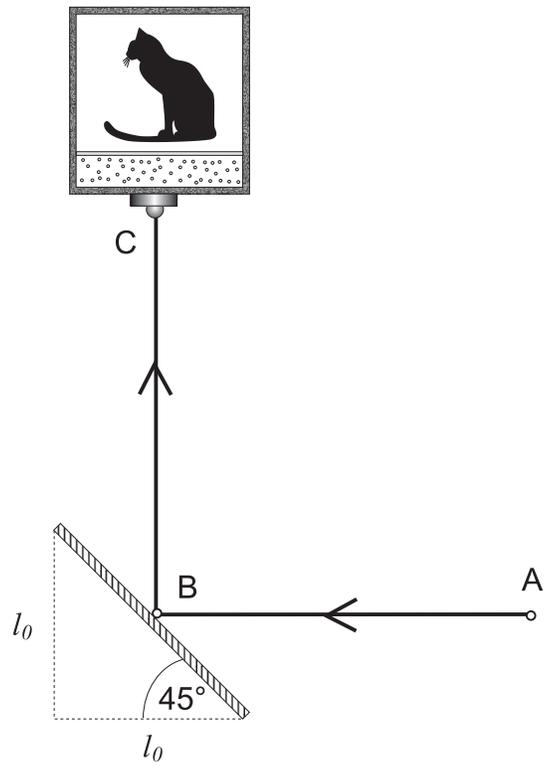

GjurchinovskiFig9

**Fig. 9.** Einstein's cat experiment No. 2. The positions of the light source and the chamber are interchanged.

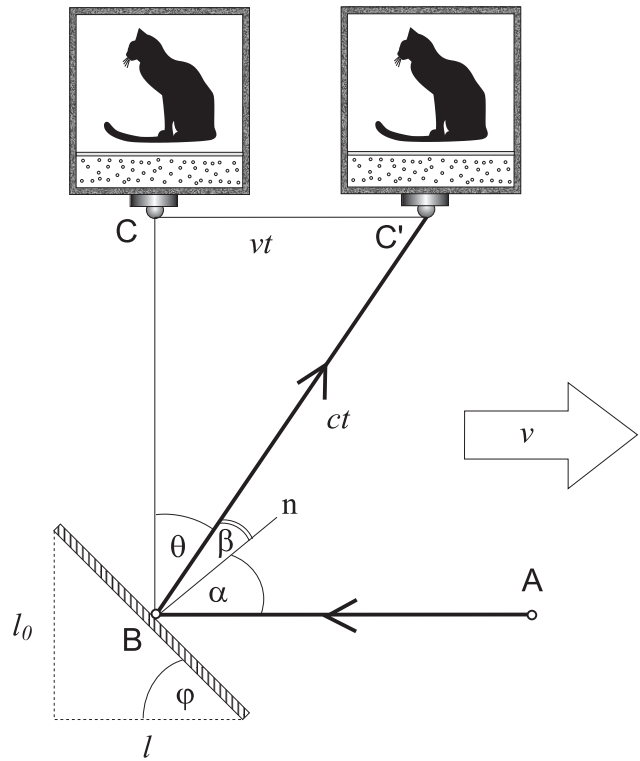

GjurchinovskiFig10

**Fig. 10.** The experiment shown in Fig. 9, observed in a reference frame where the setup is in uniform rectilinear motion.

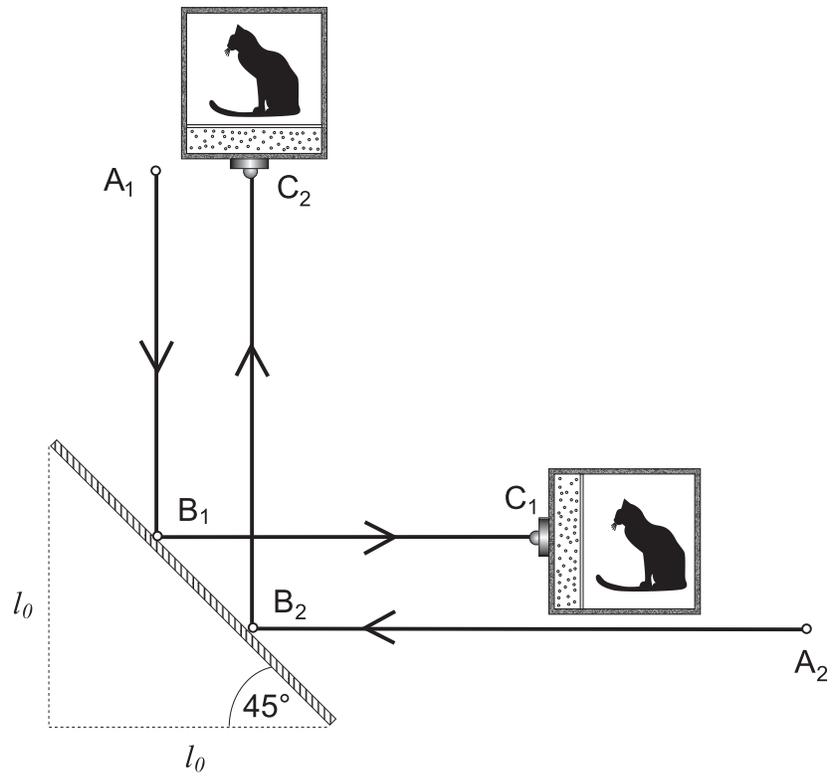

GjurchinovskiFig11

**Fig. 11.** Modified version of the Einstein's cat experiment.

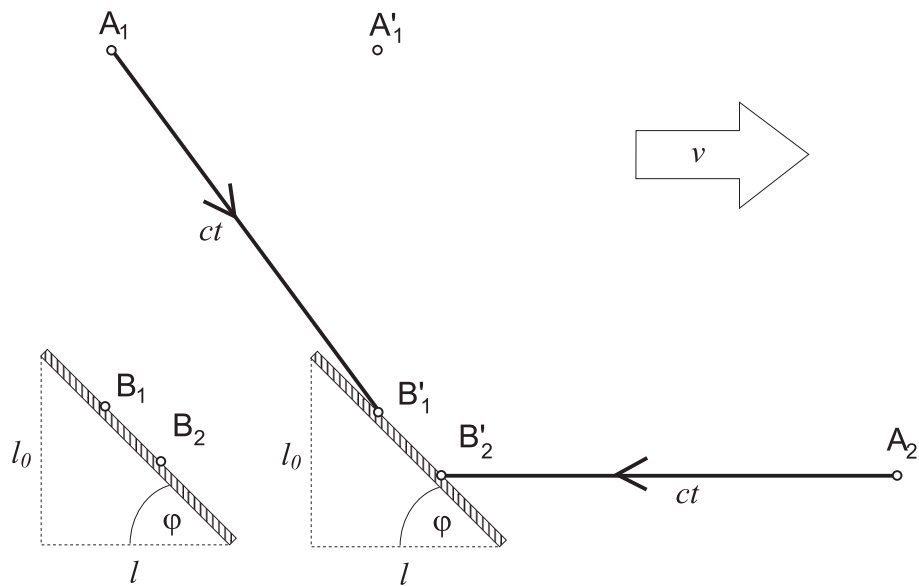

GjurchinovskiFig12

**Fig. 12.** In the reference frame where the setup is in uniform rectilinear motion, the wavefronts emitted simultaneously from the light sources $A_1$ and $A_2$ will hit the moving mirror at the same instant of time.

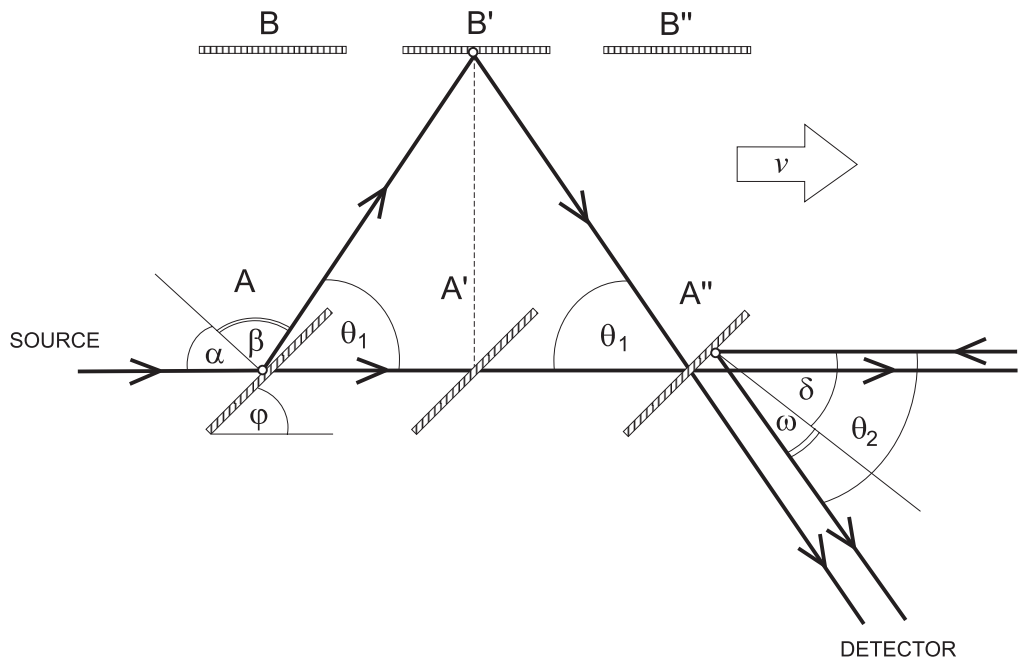

**Fig. 13.** The setup of Michelson and Morley in uniform rectilinear motion.

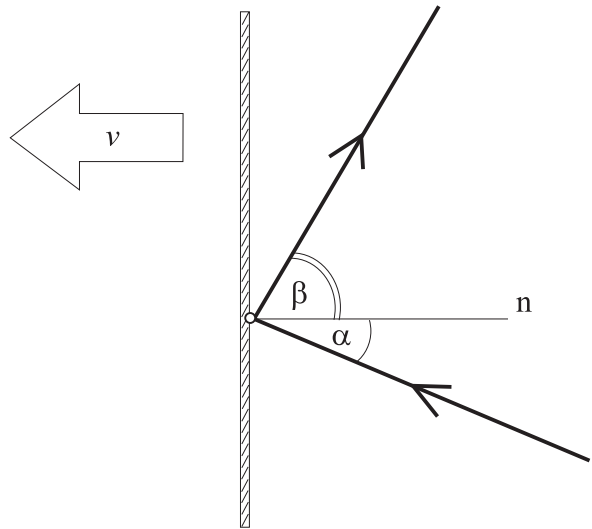

GjurchinovskiFig14

**Fig. 14.** Reflection of an electromagnetic wave by a vertical mirror moving at constant velocity *v* to the left.